# Cybersecurity of Renewable Energy Data and Applications Using Distributed Ledger Technology

Umit Cali, *Senior Member, IEEE,* Murat Kuzlu, *Senior Member, IEEE,* Manisa Pipattanasomporn, *Senior Member, IEEE,* Onur Elma, *Member, IEEE,* and Ramesh Reddi

*Abstract*— Renewable energy sources (RES) are among the most popular emerging energy resources during the past two decades. Many countries have introduced various energy policy instruments, such as renewable energy certificates (RECs), to support the growth of RES. RECs are tradable non-tangible assets, which have a monetary value. Tracking and certification of the origin of an energy resource regardless of its type (e.g., a conventional power plant or RES) is a critical operation. In addition to the certification of origin, trading transactions are needed to be performed using a secure method. Energy industry participants need to secure the data and applications related to RECs. Distributed ledger technology (DLT) is a perfect framework that can support such REC functionalities. This paper addresses the cybersecurity aspects in REC trading using Blockchain and a distributed ledger technology, considering detailed cybersecurity perspectives.

## I. Introduction

Decentralization, decarbonization, and digitalization are the three most popular trends in the energy sector. Renewable energy sources (RES), such as wind and solar generators, have been among the most well-known energy generation technologies during the past two decades. RES can be connected to power grids at the transmission, distribution, and customer levels. Large-scale RESs are typically connected at the transmission level, while small-scale and Behind-the-Meter (BTM) RESs are connected at the distribution/customer systems. Although the primary revenue source of RES is to sell clean electricity, other derivative products, which can provide additional benefits to project owners, are such as renewable energy tax credits and renewable energy certificate (REC) trading. Currently, participation in renewable energy markets may require certificates-of-origin. These are, such as GOs in the EU, RECs in the U.S., and I-RECs globally [1]. A blockchain-based REC trading platform that is owned and managed by a consortium of market participants is under development by Power Ledger, together with Midwest Renewable Energy Tracking System (M-RETS) and Clearway Energy Group in the U.S. [2].

U, Cali is with Department of Electric Power Engineering, Norwegian University of Science and Technology, Trondheim, Norway, e-mail: umit.cali@ntnu.no.
M. Kuzlu is with Department of Engineering Technology, Old Dominion University, Norfolk, VA, USA, e-mail: mkuzlu@odu.edu.
M. Pipattanasomporn with Smart Grid Research Unit, Chulalongkorn University, Bangkok, Thailand, e-mail: manisa.pip@chula.ac.th.
R. Reddi with Cyber Security Consultant, Raleigh, NC, USA, e-mail: ramesh.m.reddi@gmail.com.

The study in [3] investigated renewable energy policy mechanisms for transforming the electricity industry into a low carbon market, focusing on an auction-based mechanism and the trade-off between direct policy and integration costs. According to this study, auctions are certainly better than REC in terms of costs. These certificates provide detailed proof of origin for each megawatt-hour (MWh) of renewable generation. Blockchain-based applications are being developed to manage these systems, such as to document the provenance of renewable energy generation, issue a certificate about the green attributes of each unit of renewable generation, track the ownership and perform compliance tasks. Blockchain technology has been deployed in many applications in the electric power sector. These applications include peer-to-peer (P2P) electricity transactions, energy financing, sustainability attributions, and electric vehicle charging/billing [4]. NIST has reviewed the latest blockchain technologies recently [5], [6].

The Energy Web Foundation has organized an effort to create open-source platforms to deploy blockchain applications in the energy sector [7], [8]. Authors in [9] are involved in defining cybersecurity standards for blockchain applications for energy and utilities. Authors in [10] propose a conceptual architecture for an exchange of solar electricity in a neighborhood using the Hyperledger blockchain framework. In the other study [11], a blockchain-based conceptual framework has been proposed, which enables a home to autonomously exchange electricity in a neighborhood to locally balance renewable energy production. Authors in [12] propose a blockchain-based P2P trading scheme that couples energy and carbon markets. This can guarantee the security and privacy of the trading platform, as well as solve the inappropriate market clearing price and carbon reduction imbalance. A P2P blockchain-based energy trading platform has been proposed for residential communities in [13] to reduce the peak demand and electricity bills. This paper addresses the cybersecurity aspects involved in the blockchain-based framework for renewable energy certificates (RECs) of origin and distributed energy resources. Key cybersecurity functions are discussed, and related methodologies are described. A use case is also presented that handles cybersecurity functions to secure REC data and applications using the National Institute of Standards and Technology (NIST) Smart Grid Cybersecurity Controls.

## II. ENERGY-RELATED DLT USE CASES AND REC TRADING

As energy systems are becoming more decentralized, distributed, and intelligent, this paves the way for the next generation of digitalization technologies, such as DLT [19]. According to [20] and [21], the most popular energy-related DLT use cases are the followings:

- Peer-to-peer (P2P) electricity trading
- EV charging and payment settlement
- Labeling and energy provenance
- Grid and flexibility management
- Wholesale and retail trading (early stages)
- REC trading

P2P electricity trading is the most commonly known energy-related DLT use case, where excess electricity generated in a local market is traded in a neighborhood, and in some cases, between neighborhoods.

Fig. 1 demonstrates the segmentation of various energy-related DLT use cases on the electrical energy value chain, starting from large-scale power generation/transmission/distribution systems, retailers/ aggregators to prosumers/consumers. A prosumer is a new type of participant in a power market, acting as both a consumer and a producer of electrical energy. This is such a house with PV panels. In addition to RES, like solar PV, other decentralized components, such as electric vehicles (EVs) and electrical energy storage (EES) units, can be integrated into the overall system.

Financial (monetary), data, and commodity (electricity) transfers are three kinds of transaction types that can be observed in various energy-related DLT use cases. For instance, P2P energy trading is operational under retailer/aggregator and prosumer/consumer domains (local and regional power markets). These are the segments where (1) data, (2) financial, and (3) power transactions occur. On the other hand, the REC trading use case may only accommodate (1) data and (2) financial transactions.

1. Renewable Energy Certificates (RECs) are tradable green energy certificates, which are used to prove the origin of renewable energy sources (RES), such as wind and solar. Once RES originated electrons merge into a public utility system where generation from other energy resources is also mixed, it is impossible to track each electron whether or not it is from RES. The best way to track and record these electrons is at the origination point. RECs can be sold or traded, acting as legal tracking, labeling, and accounting mechanism. Buying RECs is supporting RES and clean energy resources, which reduces greenhouse gas emissions. Each REC gets a unique tracking ID by a certifying authority to make sure that it is not double-counted. REC can then be traded on an open market. If a REC is used by the owner, it is retired and cannot be used anymore. There are basically three ways to retire the REC:

2. RECs are used by an end-user (prosumer/consumer), marketer, generator, or utility based on statutory or regulatory requirements.

3. The RECs associated with a public claim are purchased by an end-user.

4. Any component attributes of a REC is purchased for any purpose.

When a REC is retired, it cannot be sold, donated, or transferred to any other party. The owner may only make claims associated with retired RECs.

- RECs are associated with Renewable Energy Portfolio Standards, which are state laws that require utilities to generate a specific percentage of their energy mix from RES. Each REC represents the environmental benefits of 1 MWh, which is generated by RES. RECs contain data, such as project name, certificate data, certificate type, certificate tracking ID, and renewable energy type [14]. The following steps are currently essential to perform conventional REC trading:

- RES generates electricity and sends the data to a REC tracking system.
- Once a REC is validated, the origin of electricity is recorded to the system.
- Brokers collect smaller RECs and form larger REC blocks.
- RECs are forwarded to the buyers' market.
- If RECs are used by a buyer, the record regarding the energy consumption is sent to the system.
- Once RECs are used, they need to be retired.
- Optionally initial buyers may prefer to swap the REC with other buyers to generate additional transactions.
- The regulator of auditor audits annual records and provides certificates.

A DLT-based REC trading platform can increase the efficiency of the overall system and reduce the costs associated with unnecessary third parties. Transaction times can also be potentially reduced by using DLT. The following steps summarize the operational steps of the DLT-based REC trading platform:

- RES-originated electricity is labeled digitally on the spot where it is generated by using DLT. DLT-based RECs are created.
- The issued RECs were forwarded to the DLT-based REC trading platform.
- A REC tracking system records RECs.
- RECs are traded in a P2P manner.
- Once a new owner uses the REC, it is retired, and the new status of the REC is recorded in the REC tracking system so that it is not double traded.
- If the owner of a REC prefers to swap RECs to have arbitrage gain, it can be forwarded to a new owner by updating the smart contract.
- New records are kept on the REC tracking system simultaneously. In this case, REC is not retired, and it can be reused until it is used.
- Post-purchase data is sent to the Auditing/Regulatory authority, where the entire records about the REC transactions are kept and annually audited.

Fig. 2 illustrates a generic conventional renewable energy certificate market, while Fig. 3 illustrates a generic blockchain-based renewable energy certificate.

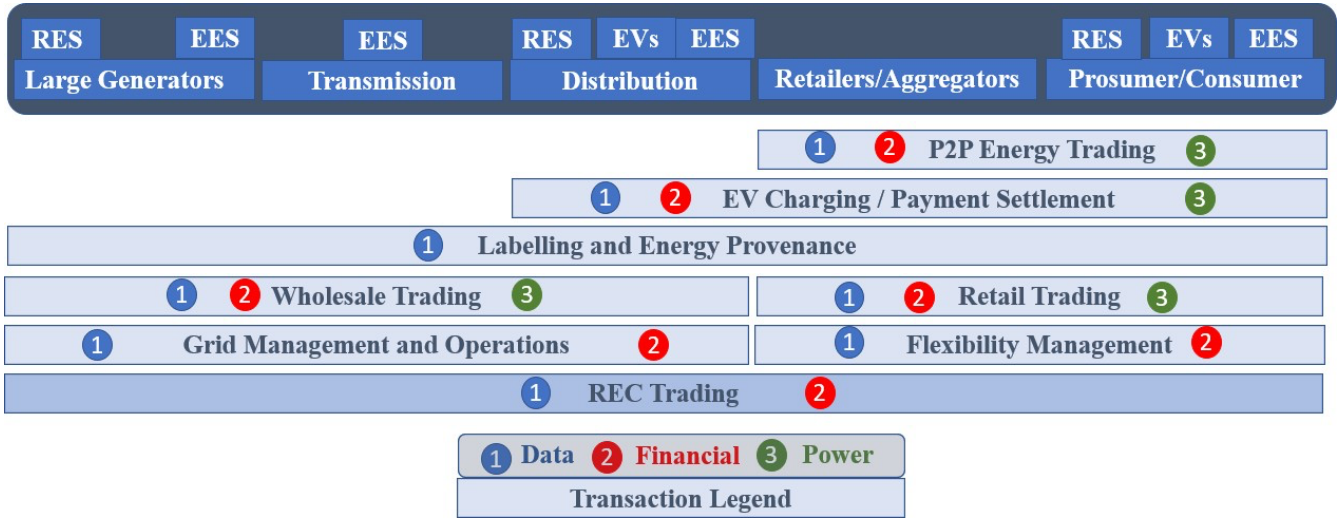

Fig. 1. Segmentation of energy-related DLT use cases.

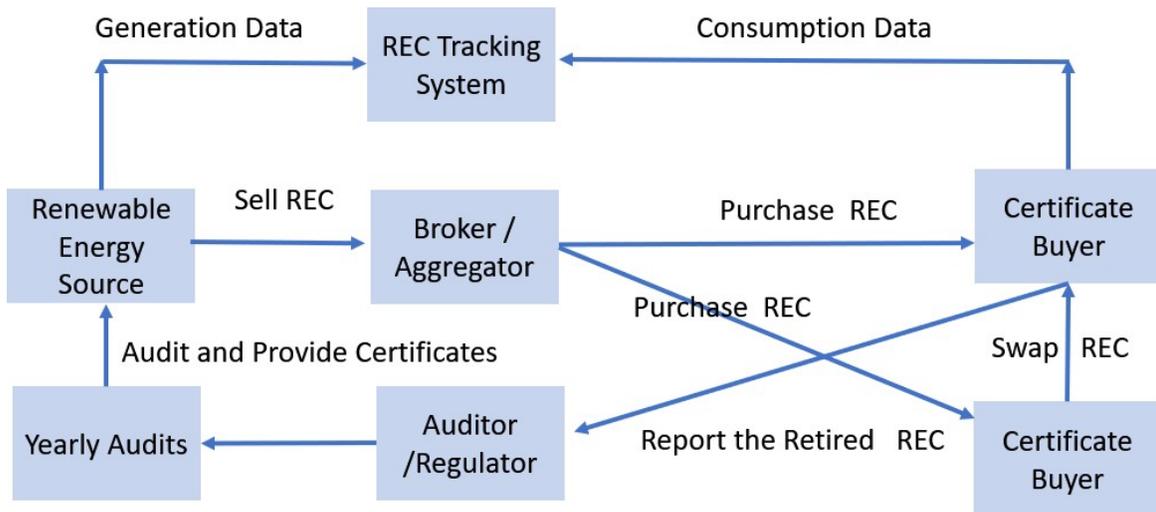

Fig. 2. Generic conventional Renewable Energy Certificate market.

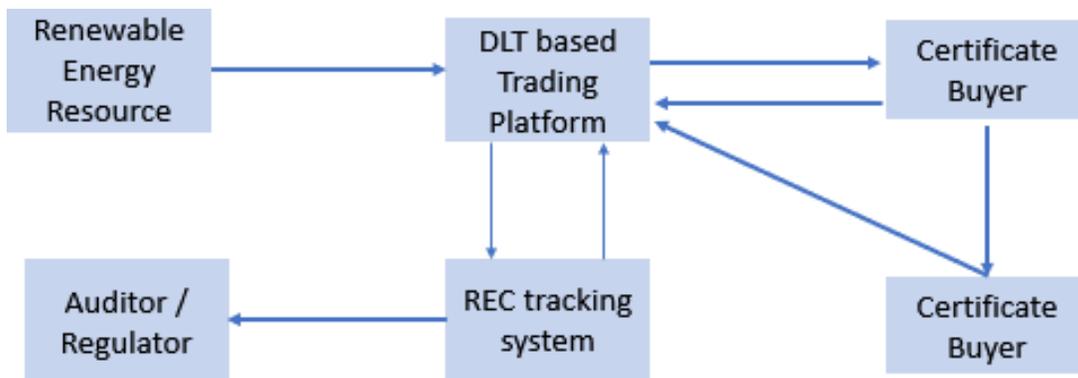

Fig. 3. Generic blockchain-based Renewable Energy Certificate.

## III. BLOCKCHAIN CAPABILITIES TO MAKE A REC SYSTEM CYBER SECURE

Cybersecurity advantages of the deployment of DLT in blockchains have been the focus of data security researchers recently [15], [16].

### A. Distributed Architecture

An advantage of the distributed architecture of the blockchain-based REC system is that it can defer or reduce the impact of a cyberattack. A REC system in a blockchain network can be permission-based to fortify further. The distributed structure of a blockchain automatically provides operational resilience as it has no single point of failure. With risk minimization through multiple nodes, an attack would not compromise the ledger stored on intact nodes. In addition, the distributed network architecture makes permissioned blockchains more secure because attackers need to compromise all or most of the nodes. However, restoring the network will experience some latency effects after an attack.

### B. Consensus Validation

A consensus mechanism is one of the main security control functions in a REC system. A consensus mechanism in a blockchain network requires a certain number of nodes to reach a consensus in order to attach a new block of data in the shared ledger. A consensus mechanism provides integrity control of the past transactions and the new blocks of data. An attempt to hack the ledger would require a consensus mechanism that can manipulate the consensus validation process to tamper with the ledger. In a permissioned blockchain network, these attacks may be prevented easily if the blockchain network contains enough nodes with a significant degree of consensus.

### C. Encryption

A permissioned blockchain network employs multiple types of encryptions to provide multilayered protection against cybersecurity threats. Asymmetric key cryptography or a public/private key encryption are widely used to meet security requirements. In addition, a combined cryptographic hashing and digital signature can be used to encrypt the linked lists or blocks.

### D. Transparency

Transparency provides another level of cybersecurity protection in a blockchain network, where each participant has an identical copy of the ledger. Several security management processes, such as real-time auditing or real-time monitoring by other participants or by a third party with limited access, can be conducted. With these security processes and proper implementation of risk management and compliance controls, identification of vulnerabilities and cybersecurity threats can be accomplished easily.

### E. Administrative Risk Controls

RES systems are hosted on systems that need to comply with regulations from authorities, such as North American Electric Reliability Corporation (NERC), and comply with standards, e.g., National Institute of Standards and Technology (NIST) Smart grid cybersecurity controls [17]. In addition, some of these systems are launched on a cloud-based platform that already implements robust cybersecurity controls across different layers of its applications. Some of the NIST cybersecurity requirements in a RES environment are discussed below. Specifically, the cases where REC system transactions involved are highlighted.

## IV. CYBERSECURITY REQUIREMENTS FOR DER ENVIRONMENTS

Based on NIST's and Electric Power Research Institute (EPRI)'s cybersecurity requirements for the DER environment, the REC process that occurs in that environment is identified through [18]. Fig. 4 illustrates the hierarchical DER system architecture. This REC process occurs at different levels. Cybersecurity requirements differ among these levels. Key levels to note are at the DER generation storage level (Level 1) and the Distribution level (Level 4).

The cybersecurity requirements for the DER environment can be described hierarchically, as follows:
- Level 1 Autonomous DER Generation and Storage
- Level 2 Facilities DER Energy Management
- Level 3 Utility and RES Operational Communications
- Level 4 Distribution Utility Operational Analysis
- Level 5 Transmission and Market Operations

*Level 1 Autonomous DER Generation and Storage* is the initial level where electricity from RES is generated and fed back to the utility grid. These systems may include solar PV, EV charging apparatus, battery storage, diesel generator, and controller. RECs need to be recorded and tracked from this origin of certification.

*Level 2 Facilities DER Energy Management* is the next level, where a Facility DER Management System (FDEMS) manages Level 1 DER systems. From the RECs point of view, they act as aggregators of RECs. The aggregated REC information is then passed to higher levels.

*Level 3 Utility and RES Operational Communications* are used to control DER systems. The systems at this level function as REC tracking systems pictured above.

*Level 4 Distribution Utility Operational Analysis* applications manage DER systems that get instructions from the distribution level. These systems interact with both transmission and DER levels. REC tracking systems interface to this level.

*Level 5 Transmission and Market Operations* is the highest level, and key players, like large utilities, Regional Transmission Organizations (RTOs), and Independent System Operators (ISOs), exchange information with REC generating and tracking systems at lower levels. This is done through the REC trading platform that does brokerage and aggregation tasks of the REC management system.

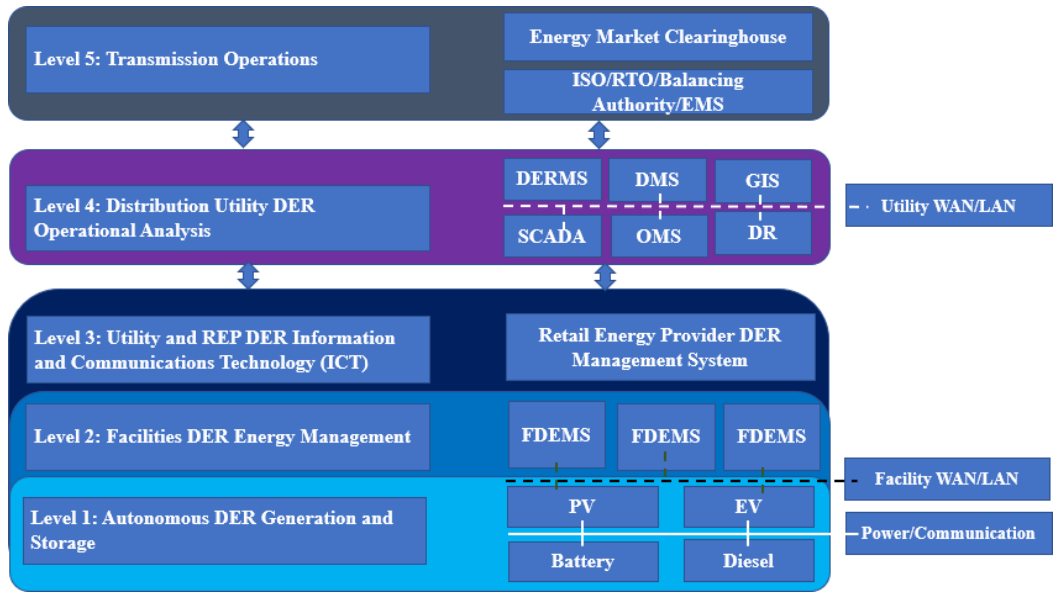

Fig. 4. The hierarchical DER system architecture

## V. Conclusion

This study investigated the cybersecurity aspects of DLT-based REC data and applications by using the commonly used conventions. Tracking, certification of the origin of the energy resource, accounting, and auditing are the main applications that manage the REC data. The cybersecurity requirements to secure the REC data and applications were mapped by using the NIST Smart Grid Cybersecurity Controls. NIST Cybersecurity Frameworks are further used in suggesting a cybersecurity maturity model for securing the REC data and applications. According to the finding of this study, DLT technology is compatible with the existing cybersecurity frameworks, but the adjustment of the overall system shall be mapped very precisely if the fully functional DLT-based REC trading platform is to be deployed on the existing power market and system. It is recommended to map the proposed use case at a higher resolution by considering the corresponding power system, communication, and cybersecurity standardization frameworks before running operational tests.

## References

[1] ECOHZ, "Guarantees Of Origin (GOS)," [Online]. Available: https://www.ecohz.com/renewable-energy-solutions/guarantees-of-origin/.

[2] American Public Power Association, "Blockchain REC trading platform set to launch in the U.S.," [Online]. Available: https://www.publicpower.org/periodical/article/blockchain-rec-trading-platform-set-launch-us.

[3] I. MacGill, A. Bruce, and S. Young, "Renewable energy auctions versus Green Certificate Schemes — lower prices but greater integration costs?," 2019 IEEE Power and Energy Society General Meeting (PESGM), Atlanta, GA, USA, 2019, pp. 1-5.

[4] David L et al., "Applying Blockchain Technology to Electric Power Systems," Council on Foreign Relations Discussion paper, Jul 2018.

[5] Dylan Y et al., "Blockchain Technology Overview," NISTIR-8202, Oct 2018.

[6] Elain B et al., "Guideline for Using Cryptographic Standards in the Federal Government: Cryptographic Mechanisms," NIST SP 800-175B, Aug 2016.

[7] Sam H et al., "Energy Web Chain: Accelerating the Energy Transition with an Open-Source Decentralized Blockchain Platform," Energy Web Foundation Publication, Oct 2018.

[8] Peter B et al., "Concept Brief: The Decentralized Autonomous Area Agent (D3A) Market Model", Energy Web Foundation Publication, 2018.

[9] Claudio Lima, "Developing Open and Interoperable DLT/Blockchain Standards," IEEE Blockchain Standards Working Group, Computer (IEEE Computer Society), Jan 2019.

[10] M. Pipattanasomporn, S. Rahman and M. Kuzlu, "Blockchain-based Solar Electricity Exchange: Conceptual Architecture and Laboratory Setup," 2019 IEEE Power and Energy Society Innovative Smart Grid Technologies Conference (ISGT), Washington, DC, USA, 2019, pp. 1-5.

[11] P. Xie et al., "Conceptual Framework of Blockchain-based Electricity Trading for Neighborhood Renewable Energy," 2018 2nd IEEE Conference on Energy Internet and Energy System Integration (EI2), Beijing, 2018, pp. 1-5.

[12] W. Hua and H. Sun, "A Blockchain-Based Peer-to-Peer Trading Scheme Coupling Energy and Carbon Markets," 2019 International Conference on Smart Energy Systems and Technologies (SEST), Porto, Portugal, 2019, pp. 1-6.

[13] S. Saxena, H. Farag, A. Brookson, H. Turesson, and H. Kim, "Design and Field Implementation of Blockchain Based Renewable Energy Trading in Residential Communities," 2019 2nd International Conference on Smart Grid and Renewable Energy (SGRE), Doha, Qatar, 2019, pp. 1-6.

[14] Federal Renewable Energy Certificate (FERC), "Guide Office of Federal Sustainability Council on Environmental Quality," June 2016.

[15] Iuon-Chang Lin and Tzu-Chun Liao, "A Survey of Blockchain Security Issues and Challenges," International Journal of Network Security, Vol 19, No 5, PP 653-659, Sep 2017.

[16] Enisa, "Distributed Ledger Technology and Cybersecurity," Dec 2016.

[17] SGIP Cyber Security Working Group," Smart Grid Cyber Security: Vol 1-3, NISTIR-7628, Aug 2010.

[18] A. Lee et al., "Cybersecurity for DER systems," EPRI/NESCOR Report 2013.

[19] Cali U, Fifield A., "Towards the decentralized revolution in energy systems using blockchain technology," International Journal of Smart Grid and Clean Energy. 2019;8(3):245-56.

[20] Cali U, Lima C. Energy Informatics Using the Distributed Ledger Technology and Advanced Data Analytics. In Cases on Green Energy and Sustainable Development 2020 (pp. 438-481).

[21] U. Cali, C. Lima, X. Li, and Y. Ogushi, "DLT/Blockchain in Transactive Energy Use Cases Segmentation and Standardization Framework," 2019 IEEE PES Transactive Energy Systems Conference (TESC), Minneapolis, MN, USA, 2019, pp. 1-5.